%% file: main.tex
\PassOptionsToPackage{table}{xcolor}
\documentclass[11pt]{article}
\usepackage[final]{acl}
\usepackage{times}
\usepackage{latexsym}
\usepackage[T1]{fontenc}
\usepackage[utf8]{inputenc}
\usepackage{booktabs}
\usepackage{graphicx}
\usepackage{adjustbox}
\usepackage{makecell}
\usepackage{algorithm}
\usepackage[noend]{algpseudocode}
\usepackage{amsmath,amssymb}
\usepackage{microtype}
\definecolor{oursrow}{HTML}{DCEBFA}   

\input{tables/numbers.tex}
\input{tables/extra_numbers.tex}
\title{eDIAR: Training-Free Open-Speaker Streaming Diarization\\for Long Meetings and Large Archives via e-means}
\author{Kyudan Jung$^{1}$ \quad Jihwan Kim$^{1}$ \quad Minwoo Lee$^{1}$ \quad Eunji Shin$^{2}$ \quad Jaegul Choo$^{1}$ \\ $^{1}$KAIST AI \quad $^{2}$Ewha W. University}

\begin{document}
\maketitle

\begin{abstract}
Meetings last up to an hour, and nobody knows in advance how many people will speak. Neural diarizers with a fixed number of output slots, such as Streaming Sortformer and Nemotron-3, are accurate and run online, but they cannot name more speakers than they have slots. We remove this limit without retraining. The slot head keeps deciding who is active and when speech overlaps, and speaker embeddings are clustered inside a single slot, with the official e-means algorithm, only after every slot has been used. In the streaming system, re-clustering is confined to a bounded window, labels are revised for a few seconds, and identities are anchored to the slots of the head.
The resulting system has no limit on the number of speakers and stays accurate when speakers outnumber the slots, where fixed-slot models break down. On meetings that grow to an hour it stays far more accurate than Streaming Sortformer, which accumulates errors, and its cost per update does not grow with the meeting. The same recipe turns the four-slot Streaming Sortformer into an open-speaker system. Across recordings, e-means makes it affordable to link the speakers of a large archive frame by frame, a task that arises when conversational corpora are prepared, and it reaches the clustering quality of Lloyd's $k$-means several times faster. The method needs little tuning, and e-means runs with its default recipe from a few hundred to more than a million embeddings.
Demo: \url{https://kyudan1.github.io/eDIAR/} \quad Model: \url{https://huggingface.co/Kyudan/eDIAR}
\end{abstract}

\section{Introduction}
Speaker diarization answers who spoke when. It is a first step of meeting transcription and is also needed to prepare multi-turn conversational audio at scale \citep{jung2026sommelier}, and a meeting assistant has to answer it while the meeting is still going on. End-to-end diarizers with a fixed number of output channels \citep{fujita2019eend} are well suited to this. They resolve the speaker permutation by arrival order and stream through a speaker cache and a FIFO queue \citep{park2024sortformer,medennikov2025streamingsortformer,medennikov2026nemotron3}. Their weakness is capacity. A four-slot model cannot output a fifth speaker, and an eight-slot model has to reuse a slot once a ninth person talks. Clustering pipelines \citep{bredin2023pyannote} have no such cap, but they decide identity from short and noisy embeddings alone, and their simple online forms commit to early decisions that they cannot repair later.

These two weaknesses are hit by different situations. Slot capacity is hit by large gatherings, whereas the drift of an online clusterer is hit by long ones, and real meetings are long. A system that is to be used for a whole meeting needs the identity stability of the slot head together with the unbounded speaker count of clustering.

We combine the two with as little machinery as possible. The slot head is trusted for activity, for overlap and, as long as a free slot exists, for identity. Clustering is used only for the one decision a fixed-slot head cannot make, namely that a single slot hosts several voices. Since a head gives a new speaker an unused slot first, we split a slot only when all slots have been used, which we call capacity-aware splitting. The splitting uses the official e-means algorithm \citep{bessa2026emeans} in its spherical form. The recipe is not tied to one model, and it turns the four-slot Streaming Sortformer into an open-speaker system as well.

The same reasoning extends beyond a single recording. Meeting series, podcast archives and corpora collected for conversational speech models contain the same people many times, and the speakers found in each session must be linked to one identity across the whole archive. That is a clustering problem with millions of embeddings and thousands of clusters, which is exactly the regime for which e-means was designed. There it matches the loss of Lloyd's algorithm in a fraction of the passes and of the time, and this makes it affordable to link every frame rather than one centroid per local speaker, the more accurate choice.

In short, a simple addition to an existing slot model keeps everything that model already does well, namely its decisions about activity and overlap, its accuracy on meetings that fit into its slots and its latency, and it adds four properties the model lacks: an unlimited number of speakers, robustness on long meetings, speaker linking across recordings and a low, constant compute cost. No retraining is needed, and the addition is a clustering step of a few hundred lines on top of the outputs of the slot model.

\begin{figure*}[t]\centering
\includegraphics[width=0.95\textwidth]{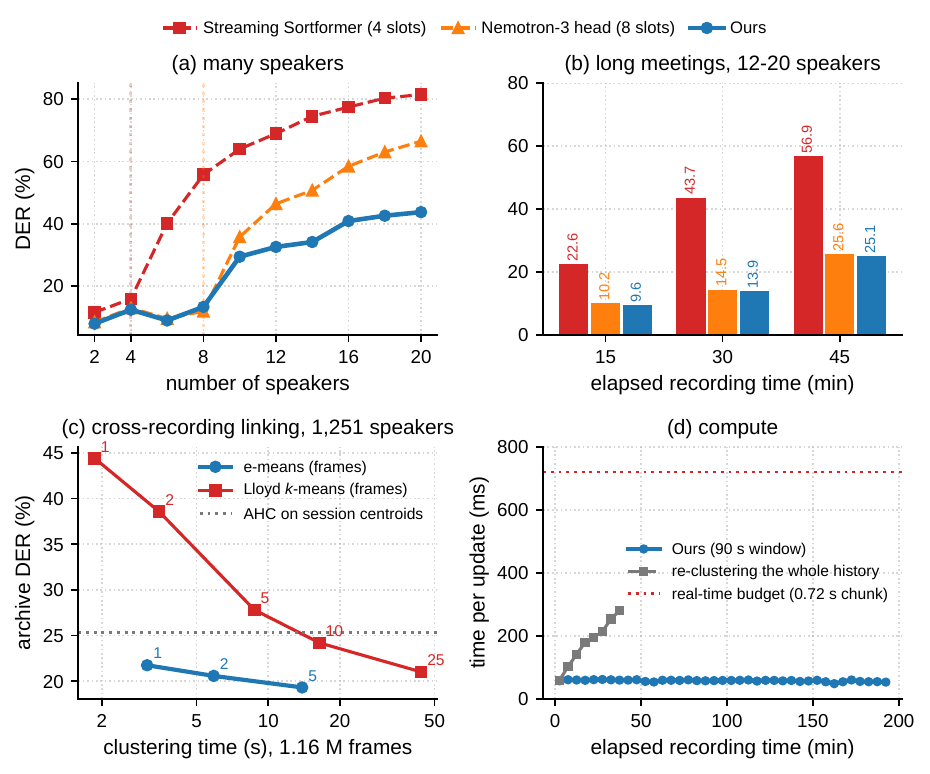}
\caption{\textbf{eDIAR at a glance.} Streaming at 1.04\,s. (a) DER against the number of speakers, from two to twenty, on LibriSpeech-OpenSpk conversations, compared with the two fixed-slot models; the dotted lines mark four and eight slots. (b) DER of six one-hour meetings with 12 to 20 speakers (LibriSpeech-OpenSpk utterances, speakers joining over time), scored on their first 15, 30 and 45 minutes. (c) Linking the speakers of 1,248 sessions with 1,251 VoxCeleb1 speakers across recordings: archive DER against clustering time at the level of 1.16 million frames; numbers next to the markers are passes over the data. (d) Processing time per update over the same long recordings, for our bounded 90\,s window and for re-clustering the whole history, with the real-time budget of one 0.72\,s chunk.}\label{fig:length}
\end{figure*}

Our system offers four practical advantages, summarised in Figure~\ref{fig:length} and demonstrated in turn.
\begin{itemize}\itemsep0pt
\item \textbf{No cap on the number of speakers.} Any slot head becomes open-speaker without retraining, shown for Nemotron-3 and Streaming Sortformer (Section~\ref{sec:nocap}).
\item \textbf{Robust as meetings grow longer.} Bounded-window re-clustering and slot-anchored identities keep the error well below that of Streaming Sortformer on hour-long meetings and ahead of the slot head (Section~\ref{sec:long}).
\item \textbf{Linking speakers across recordings.} Frame-level linking with e-means finds the same person across an archive of more than a thousand speakers (Section~\ref{sec:archive}).
\item \textbf{Low and constant compute.} The cost per update does not grow with the meeting, and e-means clusters a million frames in seconds on a CPU (Section~\ref{sec:cost}).
\end{itemize}

\section{Related Work}
\paragraph{Clustering pipelines.} Classical diarization extracts speaker embeddings from short windows and clusters them. Embeddings range from LSTM-based d-vectors \citep{wang2017speaker} to ECAPA-TDNN \citep{desplanques2020ecapa} and WeSpeaker ResNets \citep{wang2023wespeaker}, and clustering from agglomerative and spectral methods to Bayesian HMMs \citep{landini2020vbx}. pyannote~3.1 \citep{bredin2023pyannote,plaquet2023powerset} combines a segmentation model with such embeddings and clustering. Supervised alternatives replace the clustering step by a recurrent model \citep{zhang2018fully} or by a neural clusterer \citep{li2019discriminative}. \citet{park2021review} survey the field. Online variants commit to early decisions, and the classical online and mini-batch $k$-means \citep{macqueen1967some,sculley2010web,lloyd1982least,arthur2007kmeanspp} serve as our baselines.
\paragraph{End-to-end and hybrid diarization.} EEND \citep{fujita2019eend} predicts speaker activity directly, with an attractor module for a flexible number of speakers \citep{horiguchi2020eda}. Online versions use a speaker-tracing buffer \citep{xue2020online}, handle overlap and a flexible speaker count while streaming \citep{xue2021streaming}, or use global and local attractors for an unlimited number of speakers \citep{horiguchi2022online}. Hybrids combine EEND with clustering to cope with many speakers \citep{kinoshita2020integrating,kinoshita2021advances}, and target-speaker VAD refines activity given speaker profiles \citep{medennikov2020tsvad}. NeMo's multi-scale diarization decoder weights embeddings of several window scales \citep{park2022multiscale}. Our system is a hybrid in the same spirit, with one difference: clustering is confined to one slot and only after capacity is exhausted.
\paragraph{Fixed-slot neural diarization.} Sortformer \citep{park2024sortformer} orders output channels by speaker arrival time. Streaming Sortformer \citep{medennikov2025streamingsortformer} adds an Arrival-Order Speaker Cache and a FIFO queue. Nemotron-3 Diarization \citep{medennikov2026nemotron3} supports up to eight speakers and several input-buffer latencies, and we use its released streaming API as the slot head.
\paragraph{Speaker linking.} Linking the speakers of many recordings to common identities has been studied as speaker linking in large data sets \citep{vanleeuwen2010linking} and as joint diarization and linking of large corpora \citep{ferras2012linking}, usually by agglomerative clustering of one model per local speaker. We revisit it with clustering at the level of frames, which requires an engine that scales to millions of points and thousands of clusters.
\paragraph{e-means.} e-means \citep{bessa2026emeans} replaces the iterations of Lloyd's algorithm by mini-batch updates of an exponential moving average, as used for codebooks in vector-quantised autoencoders \citep{oord2017neural}, and is released in Faiss \citep{douze2024faiss}, with batch size and step size taken from fitted scaling laws. Its selling point is that it reaches a given clustering loss in far fewer passes over the data. It needs a fixed number of clusters, so choosing it is part of our contribution.

\section{Method}\label{sec:method}
\paragraph{Inputs.} At 80\,ms frame rate we have slot posteriors $p_t\in[0,1]^8$ from the Nemotron-3 head (99.2\,M parameters) and a unit-norm speaker embedding $e_t\in\mathbb{R}^{256}$ from WeSpeaker ResNet34-LM (6.6\,M parameters), pooled over a 1.52\,s window.
\paragraph{Activity.} Slot $s$ is active at $t$ if $p_{t,s}>0.55$. If no slot is active but $\max_s p_{t,s}>0.4$, the arg-max slot is activated (\emph{rescue}), provided the frame embedding has cosine $\ge0.55$ with that slot's centroid. A \emph{solo} frame has exactly one active slot.
\paragraph{Capacity-aware splitting.} A slot is \emph{used} if it is active for at least 2\,s. If fewer than all 8 slots are used, every slot is one speaker, represented by the mean direction of its solo-frame embeddings. If all slots are used, a slot may host several speakers: for $k=2,\dots,4$ we run spherical e-means on the slot's solo-frame embeddings (official implementation, default recipe: 5 epochs, batch and step from the scaling laws, EMA update with cosine-annealed step, revival, last-epoch closing means) and accept the largest $k$ for which all centroid pairs have cosine $<0.6$ and every cluster holds at least $\max(2\,\text{s},3\%$ of the slot's solo frames). Slots used for under 2\,s (but at least 0.3\,s) receive one centroid from all their frames.
\begin{algorithm}[t]
\caption{One update of the streaming system, called after every chunk of the slot head.}\label{alg:update}
\small
\begin{algorithmic}[1]
\Require slot posteriors $p_t\in[0,1]^8$ and embeddings $e_t$ of the last 90\,s; permanent ids registered per slot
\State $A_{t,s}\gets[p_{t,s}>0.55]$; if no slot is active but $\max_s p_{t,s}>0.4$, activate $\arg\max_s p_{t,s}$ when $\cos(e_t,\mu_s)\ge0.55$
\State $U\gets\{s:\text{slot } s \text{ active for at least 2\,s}\}$
\For{each slot $s$}
  \State $S_s\gets$ embeddings of the solo frames of $s$; $C_s\gets\{\text{mean direction of }S_s\}$
  \If{$|U|=8$} \Comment{capacity exhausted}
    \For{$k=4,3,2$}
      \State $\{c_1,\dots,c_k\}\gets$ spherical e-means$(S_s,k)$
      \If{$\max_{i\ne j}\cos(c_i,c_j)<0.6$ and every cluster holds $\ge\max(2\,\text{s},3\%\,|S_s|)$}
        \State $C_s\gets\{c_1,\dots,c_k\}$; \textbf{break}
      \EndIf
    \EndFor
  \EndIf
\EndFor
\State label each active slot of every frame with its closest centroid in $C_s$, smoothed over 0.4\,s
\State map the centroids of $C_s$ to the ids registered for slot $s$ (Hungarian, cosine $\ge0.4$); unmatched ones get new ids
\State emit the labels of the new frames and revise those younger than 10\,s
\end{algorithmic}
\end{algorithm}

\paragraph{Labelling.} For each active slot the frame is assigned to the best sub-cluster of that slot (cosine, smoothed over 0.4\,s); overlapping slots give overlapping labels. Clusters that cover less than 0.3\,s are re-assigned to the nearest cluster. In the non-streaming variant, clusters from different slots are merged if their centroid cosine exceeds 0.62 and the slots are rarely co-active (a cannot-link constraint), which is disabled in streaming.
Algorithm~\ref{alg:update} summarises one update of the streaming system. All thresholds were set on the \emph{dev} half (Section~\ref{sec:setup}).

\section{Streaming System}\label{sec:stream}
\paragraph{Slot head.} The Nemotron-3 streaming API is called chunk by chunk and carries its own speaker cache and FIFO. We use its low-latency mode with a 1.04\,s input buffer and its very-low-latency mode with 0.64\,s.
\paragraph{Causal embeddings.} Mel features are normalised with a cumulative mean, and the ResNet runs on short blocks with a little history and look-ahead. A causal embedding pools the window that ends at the current frame and is available immediately. The centred embedding, which is the original pooling window, becomes available 0.72\,s later and replaces the causal one when labels are recomputed. This retroactive correction matters when speakers change quickly.
\paragraph{Online clustering with revision.} After each chunk, the slot-conditioned clustering is re-run on the last 90\,s. Labels become visible as soon as they are computed (\emph{as first emitted}) and may be revised by later re-clustering until they are 10\,s old; older labels are frozen.
\paragraph{Slot-anchored speaker ids.} A permanent id is tied to the slot, because the speaker cache keeps a slot bound to one speaker over time. An unsplit slot keeps its id. The sub-clusters of a split slot are matched to the ids already registered for that slot by centroid cosine with the Hungarian algorithm, and unmatched ones receive new ids. Once all slots have been used, splitting stays enabled for the rest of the session. Anchoring ids to slots keeps them stable over long meetings.

\section{Experimental Setup}\label{sec:setup}
\paragraph{Data.} VoxConverse \citep{chung2020voxconverse} with 44 sessions, 15 with 1--4, 15 with 5--8 and 14 with 9+ speakers, cut to the first 480\,s; AMI \citep{carletta2005ami} \emph{ihm-mix} test with 16 meetings of 420\,s; ICSI \citep{janin2003icsi} test with 15 meetings of 420\,s; and LibriSpeech-OpenSpk, 18 synthetic conversations of $K\in\{2,4,6,8,10,12\}$ speakers, three each, built from LibriSpeech \citep{panayotov2015librispeech} test-clean with 0.12--0.28\,s pauses and a 0.35\,s overlap every seventh turn. Within every corpus, alternate sessions form a \emph{dev} half used for all tuning and a \emph{held-out} half. Appendix~\ref{app:data} describes every evaluation set and how it was built.
\paragraph{Metrics.} DER with a 0.25\,s collar and overlap scored with pyannote.metrics; hypotheses are merged across gaps shorter than 0.16\,s. Speaker-count accuracy, SCA, is the share of sessions with the exact number of speakers, and MAE is the mean absolute error of the count. cpWER is the concatenated-permutation WER obtained when the \emph{reference} words are tagged by the system's speakers, using oracle transcripts, so it isolates diarization errors.
\paragraph{Systems.} We compare against the official Streaming Sortformer 4spk v2.1 \citep{medennikov2025streamingsortformer} in its low-latency and its 30.4\,s configuration, the Nemotron-3 head alone, and pyannote~3.1 with default settings. Three clustering baselines operate on the same embeddings as our system, concatenated with the slot posteriors, behind a simple activity gate: leader--follower, mini-batch $k$-means and Lloyd $k$-means++. Leader--follower \citep{macqueen1967some} processes embeddings in time order, assigns each to the nearest cluster when the cosine similarity is high enough, otherwise opens a new cluster, updates the cluster mean, and never revises an earlier label. Its results are reported in Appendix~\ref{app:lf}. In the non-streaming setting the head runs in its 30.4\,s configuration and clustering sees all embeddings of the session. In the streaming setting audio is fed in 100\,ms chunks, as in the live server, and clustering baselines get streaming embeddings. Inference runs on one NVIDIA RTX~3090 with 24\,GiB of a shared server with two Intel Xeon Gold 6230 CPUs of 80 threads, and all clustering runs on the CPU.

\section{Results}\label{sec:results}
\begin{table*}[t]\centering\small
\input{tables/tab_ns}
\caption{\textbf{Non-streaming.} Mean over all sessions of each corpus, DER and SCA in \%. Best per column in bold, second best underlined. Streaming Sortformer is a four-speaker model and cannot output more than four speakers.}\label{tab:ns}
\end{table*}

\begin{table*}[t]\centering\small
\input{tables/tab_st}
\caption{\textbf{Streaming.} Same protocol as Table~\ref{tab:ns}. ``As first emitted'': labels as shown in real time; ``after revision'': labels once they are 10\,s old. Bold and underline mark the best and second best value within each latency block, 1.04\,s and 0.64\,s. Streaming Sortformer appears only in the 1.04\,s block because its model card gives just two configurations, 1.04\,s and 30.4\,s, and we did not run it at 0.64\,s. The Nemotron-3 head uses the same streaming inputs as our system. Clustering baselines are reported in Appendix~\ref{app:lf}.}\label{tab:st}
\end{table*}

\subsection{No cap on the number of speakers}\label{sec:nocap}
Tables~\ref{tab:ns} and~\ref{tab:st} compare the systems on all four corpora. Our system gives the lowest error on VoxConverse in both settings, on ICSI in both settings and on LibriSpeech in the non-streaming setting, and it estimates the number of speakers correctly in more sessions than any other system on almost every corpus. The pattern behind these numbers is simple. Wherever a session has more speakers than the head has slots, the fixed-slot models must merge people, and our system does not. Where meetings have few speakers, as in AMI and ICSI, our clustering is rarely invoked and the system keeps the accuracy of its head, so the extra capacity comes for free. Bootstrap intervals in Table~\ref{tab:boot} confirm the gains over the head and Streaming Sortformer on corpora with many speakers.

The capacity is not a property of Nemotron-3. The method only needs per-frame slot posteriors and speaker embeddings, so we also place it on the four-slot Streaming Sortformer, see Table~\ref{tab:headswap}. On VoxConverse and LibriSpeech, where most sessions have more than four speakers, our clustering removes most of the error that the speaker limit of Sortformer causes. Any slot head can thus be made open-speaker in the same way.

\begin{table*}[t]\centering\small
\input{tables/tab_boot}
\caption{DER difference, ours after revision minus competitor, with 95\,\% paired bootstrap intervals over sessions, in percentage points; negative means ours is better. $^\ast$: interval excludes zero.}\label{tab:boot}
\end{table*}

\begin{table}[t]\centering\small
\resizebox{\columnwidth}{!}{\input{tables/tab_headswap}}
\caption{DER (\%) with two different slot heads on the same embeddings, non-streaming clustering of streaming-latency inputs. The Sortformer head is thresholded per frame, without its own post-processing.}\label{tab:headswap}
\end{table}

\subsection{Robust when speakers outnumber slots}\label{sec:many}
Figure~\ref{fig:length}a follows the error as the number of speakers grows. Both fixed-slot models break down at their capacity, the four-slot Sortformer beyond four speakers and the eight-slot head beyond eight. Our system follows the head while free slots remain and degrades only gradually once they are used up. At twenty speakers its error is about two thirds of that of the head and about half of that of Sortformer. Appendix~\ref{app:pya} adds offline pyannote~3.1. The ablation in Table~\ref{tab:abl} attributes this to the two design choices, splitting a slot and doing so only when capacity is exhausted. The same ordering holds when much of the speech overlaps: with up to twelve speakers and forty percent overlap, our system stays ahead of both fixed-slot models beyond the slot count, see Appendix~\ref{app:ovsyn}.

\subsection{Robust as meetings grow longer}\label{sec:long}
Figure~\ref{fig:length}b scores six one-hour meetings of 12 to 20 speakers, built from LibriSpeech-OpenSpk utterances with speakers joining over time, on the first 15, 30 and 45 minutes. Streaming Sortformer collapses, from 23\,\% after 15 minutes to 57\,\% after 45. Our system and the Nemotron-3 head stay far below it, and ours is ahead of the head at every length and for all three speaker counts, by about half a point. The gain is small because within the 90\,s window rarely more than eight speakers are active, so the slots mostly suffice. Appendix~\ref{app:long} adds three-hour ICSI recordings, where our system behaves like the head, and a harder setting with short turns in which our error grows and passes that of the head after half an hour. Table~\ref{tab:full} confirms this on all real sessions at full length, up to an hour: our system is the best or within a point of the best on every corpus, which Streaming Sortformer does not achieve. Appendix~\ref{app:lf} shows how much online clustering degrades.

The robustness is structural. Re-clustering is confined to a bounded window, so an early mistake cannot spread through the whole meeting, and identities are tied to the slots of the head, which the speaker cache keeps bound to the same person. A running centroid bank without revision, by contrast, drifts, and every early mistake travels to the end of the meeting. A system that looks good on a short clip can be the wrong choice for a meeting.

\begin{table*}[t]\centering\small
\input{tables/tab_full}
\caption{Full-length, uncropped sessions, streaming at 1.04\,s. Mean session lengths are 8, 34 and 61 minutes.}\label{tab:full}
\end{table*}

\subsection{Linking speakers across recordings}\label{sec:archive}
A series of meetings, a podcast archive or a corpus prepared for training conversational speech models \citep{jung2026sommelier} contains the same people many times, and the speakers found in each session have to be linked to identities that hold across the whole archive \citep{vanleeuwen2010linking,ferras2012linking}. This is a hard task for two reasons. The same voice has to be recognised across changes of channel, room and microphone, and the clustering problem is large, with hundreds of thousands to millions of embeddings and as many clusters as there are people. The second is exactly the regime e-means was designed for.

\paragraph{Setup.} We built an archive from VoxCeleb1 \citep{nagrani2017voxceleb}, whose speakers do not occur in the VoxCeleb2 data on which the embedding model was trained \citep{chung2018voxceleb2}. Every speaker takes part in six sessions of six or seven people. Within a session all turns of a speaker come from one video, and the sessions of a speaker use different videos. The archive has 1,248 sessions, 1,251 speakers and 29 hours of audio. Each session is first diarized by our system. The local speakers are then linked by clustering their embeddings with $k$ set to the number of archive speakers, and every local speaker takes the archive cluster into which most of its embeddings fall. We cluster either one centroid per local speaker, segments of about one second of solo speech, or single 80\,ms frames. The archive DER is the frame-level DER over all sessions under one global mapping between hypothesis and reference identities, so a session that is diarized correctly but linked wrongly counts as an error.

\begin{figure*}[t]\centering
\includegraphics[width=0.95\textwidth]{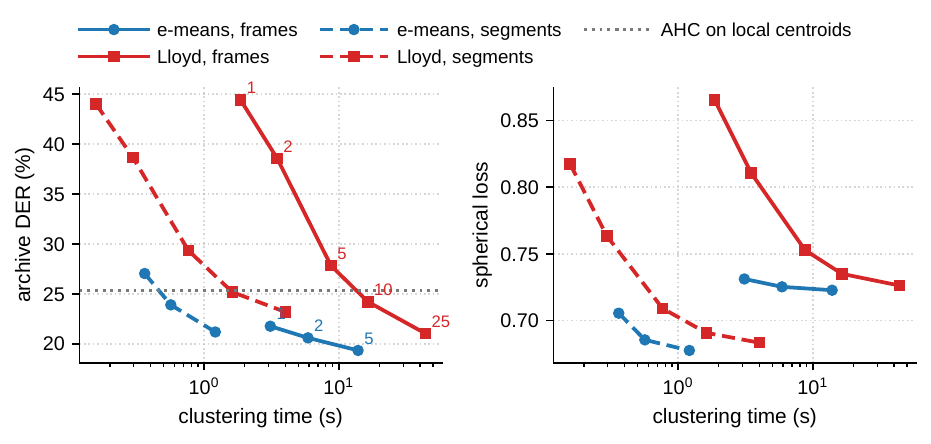}
\caption{Speaker linking across an archive of 1,248 sessions with 1,251 VoxCeleb1 speakers. Archive DER on the left and spherical clustering loss on the right against clustering time, for e-means and Lloyd's $k$-means at frame level, solid, and at segment level, dashed. Numbers next to the markers are passes over the data. The dotted line is the standard recipe, agglomerative clustering of one centroid per local speaker. CPU, 16 threads.}\label{fig:archive}
\end{figure*}

\paragraph{e-means makes frame-level linking practical.} Figure~\ref{fig:length}c, Figure~\ref{fig:archive} and Table~\ref{tab:archive} reproduce the claims of e-means on a diarization task. On frames, two epochs already reach the loss that Lloyd's algorithm reaches after twenty-five iterations, in about a seventh of its time, and on segments e-means overtakes it within a few epochs. The loss matters for the task, because a lower loss means fewer split and merged identities, and the archive DER follows it closely. Five epochs on frames give the lowest archive DER of the comparison, still in a third of the time of Lloyd. Mini-batch $k$-means, the classical way to save passes, settles at a clearly worse solution.

\paragraph{Frames beat centroids.} The standard linking recipe clusters one centroid per local speaker with agglomerative clustering. It is clearly worse than linking frames, because a centroid averages a short session and one poor centroid decides the identity of a whole local speaker, whereas frame-level linking lets every local speaker vote with all of its speech. Agglomerative clustering cannot be run on a million frames, since it needs the distances between all pairs, and Lloyd's algorithm needs many passes over them. e-means makes the more accurate frame-level solution affordable. One epoch over the whole archive takes a few seconds on a CPU, so the archive can be re-linked whenever new sessions arrive.

\paragraph{The number of clusters.} e-means needs $k$. Table~\ref{tab:archivek} sets it too low and too high. Too many clusters cost little, so in practice one can simply over-estimate the number of people, and e-means keeps its advantage over Lloyd in both directions.

\begin{table}[t]\centering\small
\adjustbox{max width=\columnwidth}{\input{tables/tab_archive}}
\caption{Speaker linking across the archive. Time is the wall-clock time of the clustering on 16 CPU threads, and loss is the spherical $k$-means objective. Rows with a second line give means and, below them, standard deviations over three seeds. $k$ is the number of speakers.}\label{tab:archive}
\end{table}

\begin{table}[t]\centering\small
\adjustbox{max width=\columnwidth}{\input{tables/tab_archive_k}}
\caption{Archive DER (\%) when $k$ is set below or above the true number of archive speakers.}\label{tab:archivek}
\end{table}

\subsection{Low and constant compute}\label{sec:cost}
The streaming system does the same amount of work at every update, because it only re-clusters a bounded window. Figure~\ref{fig:length}d shows that the time per update stays flat over the three-hour recordings, whereas re-clustering the whole history grows steadily toward the real-time budget. The per-session clustering problems are small, a few hundred embeddings and at most four clusters, so the clustering step costs milliseconds. Across an archive, e-means clusters a million frames in seconds.

\input{tables/cost_text}

\section{Conclusion}
Fixed-slot diarizers cannot name more speakers than they have slots, and online clustering drifts as a meeting grows. Without retraining, we let the slot head decide activity and overlap, split a slot only once every slot is in use, and bound error and cost over hours with a fixed window and slot-anchored identities. The result has no speaker cap, carries over to Streaming Sortformer, and, with e-means, links an archive's speakers frame by frame several times faster than Lloyd's algorithm. It keeps the accuracy of the head on meetings that fit into the slots and, at twenty speakers, lowers the error from 66 to 44\,\% for Nemotron-3 and from 81 to 44\,\% for Sortformer. The gain on long meetings is modest, and many speakers in rapid turns remain hard.

\section{Limitations}
\begin{itemize}\itemsep0pt
\item \textbf{Data.} The real corpora are small, and the open-speaker and archive benchmarks are built from read or interview speech, so they complement but do not replace large real meeting collections.
\item \textbf{Tuning.} The thresholds were set on the dev half of the same corpora, and the held-out half serves as a check.
\item \textbf{Long, speaker-rich meetings.} When more than a few speakers talk in rapid alternation for an hour, the error of our system grows with the length and can exceed that of the slot head (Appendix~\ref{app:long}).
\item \textbf{Archive linking.} The number of archive speakers is given to the clustering.
\item \textbf{Scope.} All recordings are in English, and we did not evaluate far-field rooms.
\end{itemize}

\section*{Ethics}
The live demo processes microphone audio in server memory only and stores nothing; it runs one session at a time. Speaker ids are anonymous and are not linked to identities. Linking speakers across an archive can re-identify a person across recordings. We use it only on a public research corpus, and any deployment on real recordings needs the consent of the people recorded.

\section*{Reproducibility}
All numbers in the tables are generated from result files by the scripts in \texttt{paper/}; the code, the vendored official e-means, and the streaming server are in the project repository (\texttt{AGENT.md} documents the environment).

\bibliography{references}

\appendix
\section{Evaluation Data}\label{app:data}
Table~\ref{tab:data} summarises the evaluation sets. All audio is resampled to 16\,kHz mono. Within every real corpus, alternate sessions form a dev half, which was used for all tuning, and a held-out half.

\paragraph{VoxConverse.} VoxConverse \citep{chung2020voxconverse} consists of multi-speaker clips of YouTube videos such as news, debates and talk shows, with speakers who often talk over background music or each other. We take the recordings of the first shard of the development set and the first two shards of the test set in the release on the Hugging Face Hub, group them by the number of speakers in the full recording, and keep, in the order of the release, the first fifteen recordings with one to four speakers, the first fifteen with five to eight and all fourteen with nine or more speakers in these shards, 44 recordings in total. For the main tables each recording is cut to its first 480\,s, which leaves between one and seventeen speakers per session. The full-length evaluation uses the same 44 recordings uncut, with up to twenty-one speakers and up to twenty minutes.

\paragraph{AMI.} The AMI Meeting Corpus \citep{carletta2005ami} contains meetings of four participants, most of them in a design scenario in which the group plays a product-development team. We use the sixteen meetings of the standard test split in the \emph{ihm-mix} condition, a mix of the individual headset microphones, together with the word-level transcripts and speaker labels of the release on the Hugging Face Hub. One meeting has three speakers and the others four. The main tables use the first 420\,s of each meeting, and the full-length evaluation uses the complete meetings of fourteen to fifty minutes. Word timestamps from the transcripts allow cpWER with oracle transcripts.

\paragraph{ICSI.} The ICSI Meeting Corpus \citep{janin2003icsi} records natural, unscripted research meetings of a group at the International Computer Science Institute. They have more participants than AMI, more spontaneous interaction and the largest share of overlapped speech among our corpora. We use the fifteen test meetings of the release on the Hugging Face Hub, cut to 420\,s for the main tables, which leaves four to seven speakers per session, and complete for the full-length evaluation, with forty-four to eighty-six minutes and up to ten speakers. ICSI speaker identifiers are global, because the same research group meets repeatedly, which we use for the long recordings below.

\paragraph{LibriSpeech-OpenSpk.} LibriSpeech \citep{panayotov2015librispeech} is a corpus of read audiobooks for speech recognition, in which every utterance has a single speaker. It has no conversations, so we build them from the test-clean subset, which contains forty speakers who do not occur in the training subsets. For a conversation with $K$ speakers we draw $K$ distinct speakers at random among those with at least six utterances. The conversation has three rounds, and in each round every speaker takes one turn, in a fresh random order, so that each speaker talks three times and turns of different speakers alternate. A turn is one utterance of that speaker, truncated to its first 4.5\,s, and different turns of a speaker use different utterances. Consecutive turns are separated by a pause drawn uniformly between 0.12 and 0.28\,s, and every seventh turn starts 0.35\,s before the previous one ends, which creates a short overlap. The reference segments are the exact positions of the turns, so the reference is free of annotation noise. The transcript of a turn is the LibriSpeech transcript, shortened in proportion to the truncation, and its words are spread evenly over the turn to give the word timestamps that cpWER needs. The benchmark of the main tables has three conversations for each $K\in\{2,4,6,8,10,12\}$. For the evaluation on unseen data we built ten more conversations for each of these $K$ with a different random seed, and for the many-speaker evaluation five conversations for each $K\in\{14,16,18,20\}$ with a third seed. Speaker identities can repeat between these sets because test-clean has only forty speakers, but the speaker subsets, utterances and turn orders are new.

\paragraph{Overlap-controlled conversations.} For the analysis of many speakers with heavy overlap we use the same construction with five rounds and $K\in\{6,8,10,12\}$, and we let a turn start before the previous turn ends with probability $p\in\{0,0.25,0.5,0.75\}$, by 1.0 to 2.5\,s. Three conversations per combination give 48 conversations whose share of overlapped speech ranges from zero to about forty percent.

\paragraph{ICSI meeting days.} To evaluate recordings longer than any single meeting, we join three consecutive ICSI test meetings with two seconds of silence into one recording, which gives five recordings of 2.6 to 3.2 hours with nine to eleven speakers each. Because ICSI speaker identifiers are global, a person who attends several of the joined meetings keeps one identity in the reference, as in a day of back-to-back meetings of the same group.

\paragraph{VoxCeleb1 archive.} The archive for speaker linking is built from VoxCeleb1 \citep{nagrani2017voxceleb}, interviews of 1,251 celebrities extracted from YouTube videos, whose speakers are disjoint from the VoxCeleb2 data on which the embedding model was trained \citep{chung2018voxceleb2}. Every speaker takes part in six sessions. For each of six rounds the speakers are shuffled and grouped into sessions of six, with the few remaining speakers added to sessions of seven. Within a session a speaker takes three turns, all from one video, and the six sessions of a speaker use six different videos where the speaker has that many. Turns are truncated, separated and overlapped as in LibriSpeech-OpenSpk. The result is 1,248 sessions with 29 hours of audio.

\begin{table*}[!htbp]\centering\small
\begin{tabular}{lllll}
\toprule
Set & Source & Sessions & Speakers per session & Length per session \\
\midrule
VoxConverse & YouTube clips, real & 44 & 1--17 (full: 1--21) & 480\,s (full: up to 20\,min) \\
AMI & design-scenario meetings, real & 16 & 3--4 & 420\,s (full: 14--50\,min) \\
ICSI & research meetings, real & 15 & 4--7 (full: 4--10) & 420\,s (full: 44--86\,min) \\
LibriSpeech-OpenSpk & read speech, constructed & 18 & 2--12 & 24--150\,s \\
\quad unseen sessions & read speech, constructed & 60 & 2--12 & about 90\,s \\
\quad many speakers & read speech, constructed & 20 & 14--20 & about 3.6\,min \\
Overlap-controlled & read speech, constructed & 48 & 6--12 & 85--263\,s \\
ICSI meeting days & joined real meetings & 5 & 9--11 & 2.6--3.2\,h \\
VoxCeleb1 archive & interviews, constructed & 1,248 & 6--7 (1,251 in the archive) & about 83\,s \\
\bottomrule
\end{tabular}
\caption{Evaluation sets. ``Full'' refers to the uncut recordings used in the full-length evaluation.}\label{tab:data}
\end{table*}

\section{Additional Results}\label{app:more}

\subsection{Comparison with pyannote~3.1}\label{app:pya}
\begin{figure}[t]\centering
\includegraphics[width=\columnwidth]{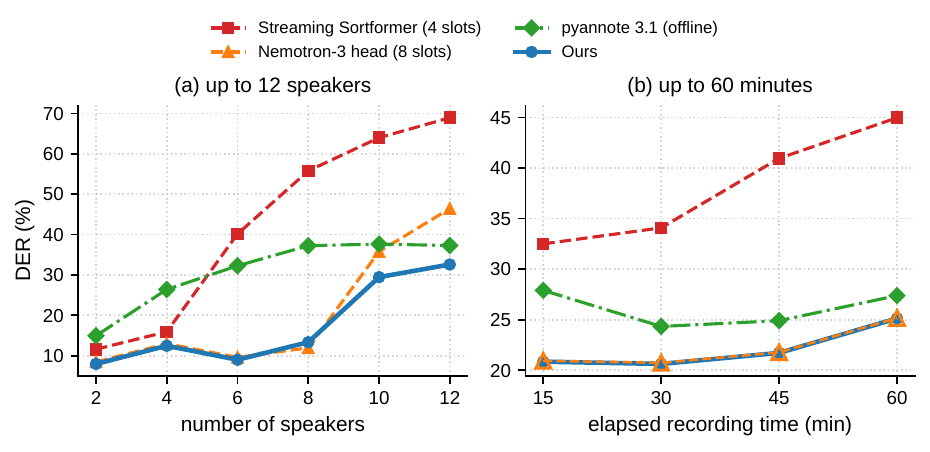}
\caption{Figure~\ref{fig:length}a and b restricted to up to 12 speakers and 60 minutes, with offline pyannote~3.1 added; hollow markers are the Nemotron-3 head.}\label{fig:pya}
\end{figure}

Figure~\ref{fig:pya} adds pyannote~3.1 with default settings to the two comparisons of Figure~\ref{fig:length}, restricted to the ranges where our system is ahead. pyannote~3.1 is an offline system: for the long recordings it is rerun from scratch on every prefix, and for the conversations it sees the whole session. Up to twelve speakers our system has a lower error than pyannote~3.1 at every speaker count (a), and on the long meetings it is lower on the first 15 to 60 minutes (b). Beyond that the picture changes: on the full 2.5\,h recordings pyannote~3.1 reaches about 29\,\% against our 31\,\%, and from sixteen speakers on its error (about 34\,\%) is below ours (41 to 44\,\%), since it is not limited by slots and sees all the audio. Our system is streaming, with a bounded cost per update.

\subsection{Longer meetings and short turns}\label{app:long}
\begin{figure}[t]\centering
\includegraphics[width=\columnwidth]{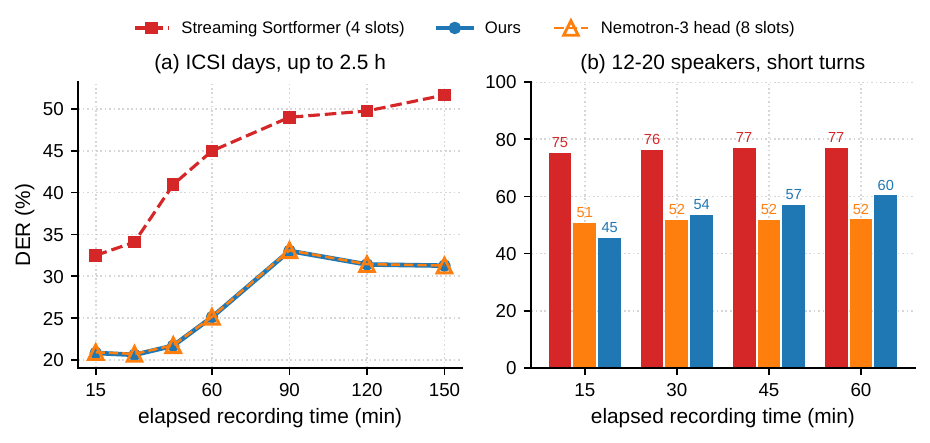}
\caption{(a) Five ICSI days of up to three hours (hollow markers: Nemotron-3 head, almost identical to ours). (b) One-hour meetings with 12 to 20 speakers, short turns, all speakers present from the start.}\label{fig:applong}
\end{figure}

Figure~\ref{fig:applong}a shows five recordings of 2.6 to 3.2 hours, each made of three consecutive ICSI meetings of the same research group and scored on their first 15 to 150 minutes. The speakers of each day fit into the eight slots, so our system is within 0.1 points of the Nemotron-3 head, and both stay far below Streaming Sortformer. Figure~\ref{fig:applong}b is a harder setting: six one-hour meetings of 12 to 20 speakers who are all present from the start and speak in short turns of 2 to 4 seconds. Our system is ahead of the head after 15 minutes (45 against 51\,\%), but its error grows with the length of the meeting and exceeds that of the head from 30 minutes on (60 against 52\,\% after an hour), while Streaming Sortformer stays at 75 to 77\,\%. We read this as a limit of the bounded window: with many speakers talking in rapid alternation the window holds only a few turns of each, and identities that the head has already assigned to slots are not corrected over time. Making long, speaker-rich meetings robust is future work.

\subsection{Online leader--follower clustering}\label{app:lf}
Table~\ref{tab:lf} reports the online leader--follower baseline next to our system in both settings, and Table~\ref{tab:bootlf} gives the paired bootstrap intervals in the streaming setting. Leader--follower processes the embeddings in time order, assigns each to the nearest cluster when the cosine similarity is at least 0.48, otherwise opens a new cluster, and never revises an earlier label. Our system estimates the number of speakers correctly far more often on the real corpora, and Figure~\ref{fig:length}b and Table~\ref{tab:fullall} show how the baseline accumulates errors on long meetings.

\begin{table*}[!htbp]\centering\small
\input{tables/tab_lf}
\caption{Online leader--follower clustering and our system, non-streaming and streaming. Bold and underline compare the two systems within each setting.}\label{tab:lf}
\vspace{6pt}
\input{tables/tab_boot_lf}
\caption{Streaming, 1.04\,s: DER difference (ours $-$ leader--follower, points) with 95\,\% paired bootstrap intervals; $^\ast$ excludes zero.}\label{tab:bootlf}
\end{table*}

\subsection{Little tuning}\label{sec:tuning}
A method that must be tuned for every domain is hard to deploy. Table~\ref{tab:hp} sweeps the main threshold of the two clustering approaches with everything else fixed, the birth threshold of leader--follower and the split threshold of our system. Leader--follower has a narrow useful range, and its best value differs between corpora, so a value that is good on one corpus costs accuracy on another. Our split threshold sits on a wide plateau. The other settings of our system are similarly flat (Table~\ref{tab:sens}). Our system with its single default setting is better on every corpus than leader--follower with the threshold chosen separately for each corpus on the evaluation data, an oracle choice that a deployed system does not have.

e-means adds no tuning of its own. Its batch size and step size follow from fitted scaling laws, and we ran its default recipe everywhere, from slot sets of a few hundred embeddings to an archive of over a million frames, without any sweep.

\begin{table}[!htbp]\centering\small
\adjustbox{max width=\columnwidth}{\input{tables/tab_hp}}
\caption{DER (\%) of the non-streaming systems when their main threshold is varied. Leader--follower is used with 0.48 elsewhere in the paper, our system with a split threshold of 0.60.}\label{tab:hp}
\end{table}

\subsection{Held-out sessions and significance}
Table~\ref{tab:held} repeats the streaming comparison on the held-out half of each corpus, which was not used for tuning, and Table~\ref{tab:nsboot} gives paired bootstrap intervals for the non-streaming system. The non-streaming system is significantly better than the Nemotron-3 head where speakers outnumber slots and than leader--follower on all four corpora.

\begin{table*}[!htbp]\centering\small
\input{tables/tab_held}
\caption{DER (\%) on the held-out half of each corpus only, that is, the alternate sessions that were not used for setting any threshold (22 VoxConverse, 8 AMI, 7 ICSI and 9 LibriSpeech-OpenSpk sessions). Upper block: streaming systems with a 1.04\,s input buffer, fed in 100\,ms chunks; ``as first emitted'' are the labels shown in real time and ``after revision'' the labels once they are 10\,s old. Lower block: non-streaming systems that see the whole session; Streaming Sortformer and the Nemotron-3 head run in their 30.4\,s configuration. Bold and underline mark the best and second best value within each block.}\label{tab:held}
\vspace{6pt}
\input{tables/tab_nsboot}
\caption{Non-streaming: DER difference (ours $-$ competitor, points) with 95\,\% paired bootstrap intervals; negative = ours better; $^\ast$ excludes zero.}\label{tab:nsboot}
\end{table*}

\subsection{Where the gain comes from}
Table~\ref{tab:abl} removes one component at a time. The gain over the head comes from splitting slots, from doing so only when capacity is exhausted, and from the low-posterior rescue.

\begin{table}[!htbp]\centering\small
\adjustbox{max width=\columnwidth}{\input{tables/tab_abl}}
\caption{Ablation of the non-streaming recipe on VoxConverse (44 sessions).}\label{tab:abl}
\end{table}

\subsection{Unseen synthetic conversations}
Table~\ref{tab:ext} evaluates 60 new LibriSpeech-OpenSpk conversations built with a seed never used before, without changing any threshold, and Table~\ref{tab:extboot} gives the bootstrap intervals. The gains over the slot head beyond its capacity reproduce on these unseen sessions, and Streaming Sortformer again fails once speakers outnumber its four slots.

\begin{table*}[!htbp]\centering\small
\input{tables/tab_ext}
\caption{Streaming systems on 60 new LibriSpeech-OpenSpk conversations (ten per $K$, unseen seed, 1.04\,s).}\label{tab:ext}
\end{table*}

\begin{table}[!htbp]\centering\small
\adjustbox{max width=\columnwidth}{\input{tables/tab_extboot}}
\caption{Bootstrap, unseen sessions; $^\ast$ excludes 0.}\label{tab:extboot}
\end{table}

\subsection{Many speakers and heavy overlap}\label{app:ovsyn}
We built 48 conversations with 6 to 12 speakers in which the share of overlapped speech is controlled and reaches about forty percent. Tables~\ref{tab:ovsyn} and~\ref{tab:ovsynk} report the streaming systems by overlap level and by number of speakers. Beyond the slot count our system is ahead of both fixed-slot models at every overlap level considered.

\begin{table}[!htbp]\centering\small
\adjustbox{max width=\columnwidth}{\input{tables/tab_ovsyn_level}}
\caption{DER (\%) on 48 synthetic conversations (6--12 speakers) by the share of overlapped speech (streaming systems, 1.04\,s).}\label{tab:ovsyn}
\end{table}

\begin{table}[!htbp]\centering\small
\adjustbox{max width=\columnwidth}{\input{tables/tab_ovsyn_k}}
\caption{The same, by number of speakers.}\label{tab:ovsynk}
\end{table}

\subsection{Overlapped speech on real meetings}
Table~\ref{tab:overlapsplit} splits the sessions of each real corpus at the median share of overlapped speech. Our clustering keeps the overlap handling of the Nemotron-3 head, since the head decides whether a second speaker is active, and it is clearly better than leader--follower on the overlap-heavy meetings. Table~\ref{tab:overlap} scores the overlapped regions alone.

\begin{table}[!htbp]\centering\small
\adjustbox{max width=\columnwidth}{\input{tables/tab_overlap_split}}
\caption{DER (\%) on the half of the sessions with the lowest and the half with the highest share of overlapped reference speech in each corpus. The second row gives the mean share of overlapped speech of each half.}\label{tab:overlapsplit}
\end{table}

\begin{table*}[!htbp]\centering\small
\input{tables/tab_overlap}
\caption{DER (\%) and missed speech (\%) inside reference-overlap regions only, collar 0, non-streaming inputs. ``overlap'' is the share of reference speech that is overlapped.}\label{tab:overlap}
\vspace{6pt}
\input{tables/tab_full_all}
\caption{Full-length, uncropped sessions, streaming at 1.04\,s, including the online leader--follower baseline. The column headers give the mean session length of each corpus.}\label{tab:fullall}
\end{table*}

\subsection{Full-length sessions, all systems}
Table~\ref{tab:fullall} repeats the full-length evaluation of Table~\ref{tab:full} with the online leader--follower baseline. The gap between leader--follower and the slot-based systems grows with the length of the corpus, from a small difference on VoxConverse to a large one on ICSI, and the baseline never finds the right number of speakers on the meeting corpora.

\subsection{The clustering engine inside a session}
\begin{figure}[!htbp]\centering
\includegraphics[width=\columnwidth]{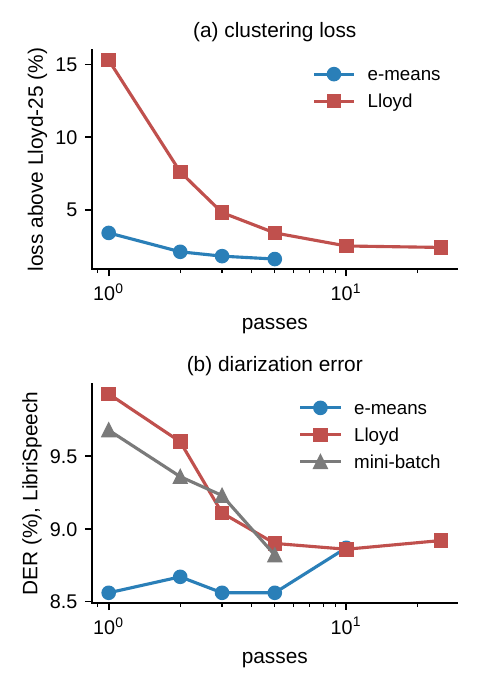}
\caption{Top: spherical clustering loss relative to 25 Lloyd iterations on the engine inputs of the benchmark. Bottom: DER of the non-streaming system on LibriSpeech-OpenSpk when the slot-splitting engine is limited to the given number of passes.}\label{fig:passes}
\end{figure}

Inside a session the splitting problems are small, and every engine gives practically the same DER, both in the non-streaming and in the streaming system (Tables~\ref{tab:engines} and~\ref{tab:engstream}). e-means reaches this DER after a single epoch, whereas Lloyd needs several iterations (Figure~\ref{fig:passes}, Tables~\ref{tab:budget} and~\ref{tab:gap}). Table~\ref{tab:scale} shows how the time of both engines grows with the problem size, which is what makes e-means the engine of choice for the archive in Section~\ref{sec:archive}.

\begin{table*}[!htbp]\centering\small
\input{tables/tab_engines}
\caption{DER (\%) of the non-streaming system with different clustering engines in the splitting step, and mean engine time per call (ms).}\label{tab:engines}
\vspace{6pt}
\input{tables/tab_scale}
\caption{Time and loss gap (relative to 25 Lloyd iterations) when real speaker embeddings are pooled to large $n$ and $k$ (faiss spherical $k$-means as Lloyd; official e-means; negative = lower loss than Lloyd-25).}\label{tab:scale}
\end{table*}

\begin{table}[!htbp]\centering\small
\adjustbox{max width=\columnwidth}{\input{tables/tab_engines_stream}}
\caption{DER (\%) after revision of the streaming system (1.04\,s) with different clustering engines in the splitting step, and the mean processing time of one update (ms, whole pipeline).}\label{tab:engstream}
\end{table}

\begin{table}[!htbp]\centering\small
\adjustbox{max width=\columnwidth}{\input{tables/tab_budget}}
\caption{DER (\%) of the non-streaming system when the slot-splitting engine is limited to a fixed number of passes (epochs, or iterations for Lloyd).}\label{tab:budget}
\end{table}

\begin{table}[!htbp]\centering\small
\adjustbox{max width=\columnwidth}{\input{tables/tab_gap}}
\caption{Spherical clustering loss relative to 25 Lloyd iterations (best of three), in \%, and time per call, on the 188 engine inputs of the benchmark.}\label{tab:gap}
\end{table}

\subsection{Sensitivity and streaming components}
Table~\ref{tab:sens} varies one hyper-parameter at a time around the default, and Table~\ref{tab:sabl} varies the streaming design. The 90\,s window balances accuracy and cost, and the length of the revision horizon changes little. Table~\ref{tab:aux} lists the speaker-count error and the oracle-transcript cpWER.

\begin{table}[!htbp]\centering\small
\adjustbox{max width=\columnwidth}{\input{tables/tab_sens}}
\caption{One-at-a-time sensitivity of the non-streaming DER to each hyper-parameter (93 sessions).}\label{tab:sens}
\end{table}

\begin{table}[!htbp]\centering\small
\adjustbox{max width=\columnwidth}{\input{tables/tab_sabl}}
\caption{Streaming ablation (93 sessions).}\label{tab:sabl}
\end{table}

\begin{table*}[!htbp]\centering\small
\input{tables/tab_aux}
\caption{Speaker-count MAE and oracle-transcript cpWER (\%); upper block non-streaming, lower block streaming (rows as in Tables~\ref{tab:ns} and~\ref{tab:st}); cpWER is only defined for AMI and LibriSpeech.}\label{tab:aux}
\end{table*}

\end{document}

%% file: tables/tab_ns.tex
\begin{tabular}{lcccccccc}
\toprule
 & \multicolumn{2}{c}{VoxConverse} & \multicolumn{2}{c}{AMI} & \multicolumn{2}{c}{ICSI} & \multicolumn{2}{c}{LibriSpeech} \\
System & DER & SCA & DER & SCA & DER & SCA & DER & SCA \\
\midrule
Streaming Sortformer 4spk, 30.4\,s config & 13.3 & 43.2 & \textbf{7.2} & \textbf{81.2} & 29.1 & 13.3 & 40.7 & 33.3 \\
Nemotron-3 (8 slots, 30.4\,s) & \underline{5.5} & \textbf{75.0} & 9.0 & \underline{68.8} & \underline{27.0} & \textbf{53.3} & 18.1 & 66.7 \\
pyannote 3.1 & 5.6 & 45.5 & 27.4 & 37.5 & 33.2 & \underline{26.7} & 32.1 & 33.3 \\
Mini-batch $k$-means (Sculley) & 7.6 & 31.8 & 13.5 & 12.5 & 36.1 & 0.0 & \underline{14.7} & 72.2 \\
Lloyd $k$-means++ (25 ep.) & 13.8 & \underline{59.1} & 9.5 & 37.5 & 36.1 & \underline{26.7} & 25.3 & \underline{77.8} \\
\rowcolor{oursrow}\textbf{Ours (non-streaming)} & \textbf{4.5} & \textbf{75.0} & \underline{8.9} & \underline{68.8} & \textbf{26.9} & \textbf{53.3} & \textbf{8.6} & \textbf{88.9} \\
\bottomrule
\end{tabular}

%% file: tables/tab_st.tex
\begin{tabular}{lcccccccc}
\toprule
 & \multicolumn{2}{c}{VoxConverse} & \multicolumn{2}{c}{AMI} & \multicolumn{2}{c}{ICSI} & \multicolumn{2}{c}{LibriSpeech} \\
System & DER & SCA & DER & SCA & DER & SCA & DER & SCA \\
\midrule
Streaming Sortformer 4spk (1.04\,s) & 14.1 & \underline{38.6} & \textbf{8.3} & \textbf{75.0} & 27.7 & 13.3 & 41.4 & 27.8 \\
Nemotron-3 head, 8 slots (1.04\,s) & 6.4 & \textbf{75.0} & 9.3 & \textbf{75.0} & \textbf{27.4} & \underline{46.7} & 19.8 & \textbf{66.7} \\
\rowcolor{oursrow}Ours, as first emitted (1.04\,s) & \underline{5.9} & \textbf{75.0} & 9.2 & \textbf{75.0} & \underline{27.5} & \textbf{53.3} & \underline{18.5} & 55.6 \\
\rowcolor{oursrow}\textbf{Ours, after $\le$10\,s revision (1.04\,s)} & \textbf{5.8} & \textbf{75.0} & \underline{9.1} & \textbf{75.0} & \textbf{27.4} & \textbf{53.3} & \textbf{16.4} & \underline{61.1} \\
\midrule
Nemotron-3 head, 8 slots (0.64\,s) & 6.8 & \underline{72.7} & \textbf{9.1} & \textbf{75.0} & \textbf{27.5} & \underline{46.7} & 20.1 & \textbf{66.7} \\
\rowcolor{oursrow}Ours, as first emitted (0.64\,s) & \underline{6.3} & \textbf{77.3} & \underline{9.2} & \textbf{75.0} & \underline{27.7} & \textbf{53.3} & \underline{19.5} & 55.6 \\
\rowcolor{oursrow}\textbf{Ours, after $\le$10\,s revision (0.64\,s)} & \textbf{6.2} & \textbf{77.3} & \textbf{9.1} & \textbf{75.0} & \textbf{27.5} & \textbf{53.3} & \textbf{16.9} & \underline{61.1} \\
\bottomrule
\end{tabular}

%% file: tables/tab_boot.tex
\begin{tabular}{lcccc}
\toprule
  & Vox & AMI & ICSI & Libri \\
\midrule
Streaming Sortformer 4spk & -8.3 [-12.3, -4.5]$^\ast$ & +0.8 [-0.9, +2.6] & -0.3 [-5.3, +4.9] & -25.0 [-32.4, -17.2]$^\ast$ \\
Nemotron-3 head (8 slots) & -0.6 [-1.5, -0.1]$^\ast$ & -0.1 [-0.4, +0.1] & -0.0 [-0.3, +0.3] & -3.3 [-5.8, -1.3]$^\ast$ \\
\bottomrule
\end{tabular}

%% file: tables/tab_headswap.tex
\begin{tabular}{lcccc}
\toprule
Slot head and clustering & Vox & AMI & ICSI & Libri \\
\midrule
Streaming Sortformer head (4 slots) & 15.8 & 10.6 & 30.8 & 41.4 \\
\rowcolor{oursrow}\quad + our clustering & 6.6 & 12.1 & 30.8 & 16.2 \\
\midrule
Nemotron-3 head (8 slots) & 6.4 & 9.7 & 26.5 & 18.1 \\
\rowcolor{oursrow}\quad + our clustering & 4.6 & 9.0 & 26.5 & 9.9 \\
\bottomrule
\end{tabular}

%% file: tables/tab_full.tex
\begin{tabular}{lcccccc}
\toprule
 & \multicolumn{2}{c}{VoxConverse} & \multicolumn{2}{c}{AMI} & \multicolumn{2}{c}{ICSI} \\
Full-length, streaming (1.04\,s) & DER & SCA & DER & SCA & DER & SCA \\
\midrule
Streaming Sortformer 4spk (1.04\,s) & 18.1 & 29.5 & 8.9 & 93.8 & 36.3 & 6.7 \\
Nemotron-3 head, 8 slots & 8.5 & 59.1 & 9.7 & 81.2 & 28.5 & 33.3 \\
\rowcolor{oursrow}Ours, as first emitted & 8.0 & 59.1 & 9.6 & 87.5 & 28.7 & 40.0 \\
\rowcolor{oursrow}Ours, after revision & 7.9 & 59.1 & 9.6 & 87.5 & 28.6 & 40.0 \\
\bottomrule
\end{tabular}

%% file: tables/tab_archive.tex
\begin{tabular}{lcccc}
\toprule
Method & Passes & Time (s) & Loss & DER (\%) \\
\midrule
\multicolumn{5}{l}{\emph{session centroids, $n=7.6$\,k}} \\
AHC (average) & -- & 5.09 & -- & 25.3 \\
e-means & 5 & 0.19 & 0.378 & 27.9 \\
Lloyd & 25 & 0.12 & 0.413 & 34.6 \\
\midrule
\multicolumn{5}{l}{\emph{1\,s segments, $n=98$\,k}} \\
e-means & 1 & \makecell{0.36\\[-1pt]{\scriptsize $\pm$0.02}} & \makecell{0.706\\[-1pt]{\scriptsize $\pm$0.001}} & \makecell{27.0\\[-1pt]{\scriptsize $\pm$0.7}} \\
e-means & 2 & \makecell{0.57\\[-1pt]{\scriptsize $\pm$0.05}} & \makecell{0.686\\[-1pt]{\scriptsize $\pm$0.001}} & \makecell{23.9\\[-1pt]{\scriptsize $\pm$0.9}} \\
e-means & 5 & \makecell{1.21\\[-1pt]{\scriptsize $\pm$0.10}} & \makecell{0.678\\[-1pt]{\scriptsize $\pm$0.001}} & \makecell{21.2\\[-1pt]{\scriptsize $\pm$0.5}} \\
Lloyd & 1 & 0.16 & 0.818 & 44.0 \\
Lloyd & 2 & 0.30 & 0.763 & 38.7 \\
Lloyd & 5 & \makecell{0.76\\[-1pt]{\scriptsize $\pm$0.04}} & \makecell{0.709\\[-1pt]{\scriptsize $\pm$0.001}} & \makecell{29.3\\[-1pt]{\scriptsize $\pm$0.2}} \\
Lloyd & 10 & 1.62 & 0.691 & 25.2 \\
Lloyd & 25 & \makecell{4.00\\[-1pt]{\scriptsize $\pm$0.13}} & \makecell{0.683\\[-1pt]{\scriptsize $\pm$0.001}} & \makecell{23.2\\[-1pt]{\scriptsize $\pm$0.4}} \\
mini-batch & 1 & 1.33 & 0.764 & 36.3 \\
mini-batch & 5 & 0.95 & 0.730 & 32.4 \\
k-means++ & 1 & 86.56 & 0.728 & 27.8 \\
k-means++ & 5 & 86.64 & 0.686 & 23.1 \\
k-means++ & 25 & 86.41 & 0.677 & 20.5 \\
\midrule
\multicolumn{5}{l}{\emph{80\,ms frames, $n=1.16$\,M}} \\
e-means & 1 & \makecell{3.10\\[-1pt]{\scriptsize $\pm$0.12}} & \makecell{0.731\\[-1pt]{\scriptsize $\pm$0.000}} & \makecell{21.8\\[-1pt]{\scriptsize $\pm$0.2}} \\
e-means & 2 & \makecell{5.89\\[-1pt]{\scriptsize $\pm$0.24}} & \makecell{0.725\\[-1pt]{\scriptsize $\pm$0.001}} & \makecell{20.6\\[-1pt]{\scriptsize $\pm$0.5}} \\
e-means & 5 & \makecell{13.88\\[-1pt]{\scriptsize $\pm$0.40}} & \makecell{0.723\\[-1pt]{\scriptsize $\pm$0.001}} & \makecell{19.3\\[-1pt]{\scriptsize $\pm$0.5}} \\
Lloyd & 1 & 1.86 & 0.865 & 44.4 \\
Lloyd & 2 & 3.48 & 0.811 & 38.6 \\
Lloyd & 5 & \makecell{8.74\\[-1pt]{\scriptsize $\pm$0.34}} & \makecell{0.753\\[-1pt]{\scriptsize $\pm$0.002}} & \makecell{27.8\\[-1pt]{\scriptsize $\pm$0.9}} \\
Lloyd & 10 & 16.42 & 0.735 & 24.2 \\
Lloyd & 25 & \makecell{43.86\\[-1pt]{\scriptsize $\pm$2.62}} & \makecell{0.726\\[-1pt]{\scriptsize $\pm$0.001}} & \makecell{21.0\\[-1pt]{\scriptsize $\pm$0.5}} \\
mini-batch & 1 & 6.18 & 0.768 & 31.0 \\
mini-batch & 5 & 27.03 & 0.757 & 29.5 \\
\bottomrule
\end{tabular}

%% file: tables/tab_archive_k.tex
\begin{tabular}{lccc}
\toprule
$k$ relative to the true number of speakers & 0.8 & 1.0 & 1.25 \\
\midrule
0.96\,s segments, e-means 2 & 32.6 & 23.9 & 22.0 \\
0.96\,s segments, Lloyd 25 & 32.0 & 23.2 & 22.1 \\
80\,ms frames, e-means 2 & 29.9 & 20.6 & 19.3 \\
80\,ms frames, Lloyd 25 & 30.3 & 21.0 & 20.8 \\
\bottomrule
\end{tabular}

%% file: tables/cost_text.tex
\paragraph{Latency.} The input buffer of Nemotron-3 is 1.04\,s in low latency mode and 0.64\,s in very low latency mode. The ResNet runs in parallel with the head and needs only 0.2 to 0.5\,s of look-ahead, which we measured by replacing the audio after a time $t_0$ with noise and finding the earliest affected frame. It therefore adds no latency. We streamed two LibriSpeech clips through the server over a TLS WebSocket at real-time speed, one with 4 speakers and 30\,s and one with 10 speakers and 60\,s. A label appeared on average 0.70\,s after the sound for 4 speakers and 0.83\,s for 10 speakers, with a 95th percentile of 1.0\,s.
\paragraph{Compute.} Nemotron-3 has 99.2\,M parameters and ResNet34 with its embedding layer has 6.6\,M. The official Streaming Sortformer 4spk checkpoint has 117.7\,M. With the PyTorch FLOP counter, counting a multiply-add as two operations, Nemotron-3 takes 45.3\,GFLOP per second of audio in low latency mode and 65.7\,GFLOP in very low latency mode. The ResNet takes 29.4\,GFLOP, because each 0.32\,s block re-computes 1.6\,s of left context and 0.16\,s of look-ahead, and the embedding layer takes 0.07\,GFLOP. In total this is about 75\,GFLOP per second of audio. We did not count the FLOPs of Streaming Sortformer.
\paragraph{GPU, CPU and memory.} With BLAS threads limited to 4, the server process held 850\,MiB of GPU memory. The average GPU utilisation was 1.4\,\% for 4 speakers and 2.1\,\% for 10 speakers, with peaks of 7 to 8\,\%. The process used on average 0.33 and 1.13 CPU cores and a constant 1.9\,GiB of resident memory. The server time per update was 56\,ms for 4 speakers, and 188\,ms on average and 346\,ms at the 95th percentile for 10 speakers. These are real-time factors of 0.074 and 0.25, since the chunk period is 720\,ms. Over all 93 benchmark sessions, fed as fast as possible on the shared host, the mean real-time factor was 0.10 with a 95th percentile of 0.24. Streaming Sortformer, run with NeMo at batch size 1 under the same conditions, had 0.07. On the longest recording, 3.2\,h of ICSI, the streaming system peaked at 0.43\,GiB of GPU memory, of which 0.40\,GiB are the weights, whereas Streaming Sortformer reached 9.6\,GiB. Both held about 3.8\,GiB of resident CPU memory. Frame-level archive linking of 1.16\,M frames with $k=1{,}251$ on 16 threads peaked at 4.5\,GiB and took 12.2\,s with e-means after 5 epochs and 46.5\,s with Lloyd's algorithm after 25 iterations.

%% file: tables/tab_lf.tex
\begin{tabular}{lcccccccc}
\toprule
 & \multicolumn{2}{c}{VoxConverse} & \multicolumn{2}{c}{AMI} & \multicolumn{2}{c}{ICSI} & \multicolumn{2}{c}{LibriSpeech} \\
System & DER & SCA & DER & SCA & DER & SCA & DER & SCA \\
\midrule
Online Leader-Follower (non-streaming) & \underline{6.2} & \underline{54.5} & \underline{9.9} & \underline{12.5} & \underline{30.2} & \underline{0.0} & \underline{10.4} & \underline{72.2} \\
\rowcolor{oursrow}\textbf{Ours (non-streaming)} & \textbf{4.5} & \textbf{75.0} & \textbf{8.9} & \textbf{68.8} & \textbf{26.9} & \textbf{53.3} & \textbf{8.6} & \textbf{88.9} \\
\midrule
Online Leader-Follower (1.04\,s) & \underline{5.9} & \underline{50.0} & \underline{10.7} & \underline{0.0} & \underline{31.6} & \underline{6.7} & \textbf{11.6} & \textbf{61.1} \\
\rowcolor{oursrow}\textbf{Ours, after $\le$10\,s revision (1.04\,s)} & \textbf{5.8} & \textbf{75.0} & \textbf{9.1} & \textbf{75.0} & \textbf{27.4} & \textbf{53.3} & \underline{16.4} & \textbf{61.1} \\
\midrule
Online Leader-Follower (0.64\,s) & \underline{6.6} & \underline{38.6} & \underline{10.8} & \underline{0.0} & \underline{31.5} & \underline{6.7} & \textbf{12.1} & \textbf{61.1} \\
\rowcolor{oursrow}\textbf{Ours, after $\le$10\,s revision (0.64\,s)} & \textbf{6.2} & \textbf{77.3} & \textbf{9.1} & \textbf{75.0} & \textbf{27.5} & \textbf{53.3} & \underline{16.9} & \textbf{61.1} \\
\bottomrule
\end{tabular}

%% file: tables/tab_boot_lf.tex
\begin{tabular}{lcccc}
\toprule
  & Vox & AMI & ICSI & Libri \\
\midrule
Online Leader-Follower & -0.1 [-1.0, +0.7] & -1.5 [-2.4, -0.8]$^\ast$ & -4.2 [-7.8, -1.9]$^\ast$ & +4.8 [+2.0, +8.1]$^\ast$ \\
\bottomrule
\end{tabular}

%% file: tables/tab_hp.tex
\begin{tabular}{lcccc}
\toprule
Setting & Vox & AMI & ICSI & Libri \\
\midrule
Leader--follower, birth threshold 0.30 & 5.1 & 10.0 & 29.1 & 13.7 \\
Leader--follower, birth threshold 0.36 & 5.1 & 9.2 & 29.4 & 10.1 \\
Leader--follower, birth threshold 0.42 & 5.2 & 9.7 & 29.3 & 9.7 \\
Leader--follower, birth threshold 0.48 & 6.2 & 9.9 & 30.2 & 10.4 \\
Leader--follower, birth threshold 0.54 & 6.4 & 11.2 & 32.5 & 12.9 \\
Leader--follower, birth threshold 0.60 & 8.0 & 12.6 & 34.6 & 16.6 \\
Leader--follower, birth threshold 0.66 & 10.8 & 13.9 & 44.8 & 20.0 \\
\midrule
Ours, split threshold 0.45 & 4.3 & 8.9 & 26.9 & 9.1 \\
Ours, split threshold 0.50 & 4.3 & 8.9 & 26.9 & 8.7 \\
Ours, split threshold 0.55 & 4.3 & 8.9 & 26.9 & 8.6 \\
Ours, split threshold 0.60 & 4.5 & 8.9 & 26.9 & 8.6 \\
Ours, split threshold 0.65 & 4.5 & 8.9 & 26.9 & 9.0 \\
Ours, split threshold 0.70 & 4.7 & 8.9 & 26.9 & 10.4 \\
Ours, split threshold 0.75 & 4.9 & 8.9 & 26.9 & 14.5 \\
\bottomrule
\end{tabular}

%% file: tables/tab_held.tex
\begin{tabular}{lcccc}
\toprule
DER (\%), held-out half & VoxConverse & AMI & ICSI & LibriSpeech \\
\midrule
\multicolumn{5}{l}{\emph{streaming, 1.04\,s input buffer}} \\
Streaming Sortformer 4spk & 14.1 & \textbf{9.0} & 23.7 & 42.5 \\
Nemotron-3 head, 8 slots & 5.6 & 10.6 & \textbf{21.7} & 21.9 \\
Online Leader-Follower & \textbf{5.2} & 12.1 & 23.7 & \textbf{11.8} \\
\rowcolor{oursrow}Ours, as first emitted & \underline{5.3} & \underline{10.4} & \underline{21.9} & 19.6 \\
\rowcolor{oursrow}\textbf{Ours, after revision} & \textbf{5.2} & \underline{10.4} & \textbf{21.7} & \underline{17.6} \\
\midrule
\multicolumn{5}{l}{\emph{non-streaming, whole session}} \\
pyannote 3.1 & \underline{5.2} & 29.0 & 27.7 & 32.5 \\
Streaming Sortformer 4spk, 30.4\,s config & 13.2 & \textbf{7.6} & 24.5 & 42.2 \\
Nemotron-3, 8 slots, 30.4\,s & \underline{5.2} & 9.9 & \textbf{20.8} & 19.5 \\
Online Leader-Follower & \underline{5.2} & 11.3 & \underline{21.7} & \underline{10.0} \\
\rowcolor{oursrow}\textbf{Ours (non-streaming)} & \textbf{4.4} & \underline{9.8} & \textbf{20.8} & \textbf{8.7} \\
\bottomrule
\end{tabular}

%% file: tables/tab_nsboot.tex
\begin{tabular}{lcccc}
\toprule
  & Vox & AMI & ICSI & Libri \\
\midrule
Streaming Sortformer 4spk (30.4\,s) & -8.8 [-12.9, -4.8]$^\ast$ & +1.7 [+0.1, +3.4]$^\ast$ & -2.2 [-5.5, +1.3] & -32.1 [-42.5, -21.3]$^\ast$ \\
Nemotron-3 (8 slots, 30.4\,s) & -1.0 [-2.3, -0.1]$^\ast$ & -0.1 [-0.1, -0.0]$^\ast$ & -0.1 [-0.3, +0.1] & -9.6 [-15.7, -4.2]$^\ast$ \\
pyannote 3.1 & -1.2 [-2.6, +0.3] & -18.5 [-22.7, -14.3]$^\ast$ & -6.3 [-9.3, -3.4]$^\ast$ & -23.5 [-31.3, -15.6]$^\ast$ \\
Online Leader-Follower & -1.8 [-4.3, -0.2]$^\ast$ & -1.0 [-1.6, -0.5]$^\ast$ & -3.3 [-7.5, -0.8]$^\ast$ & -1.9 [-3.7, -0.4]$^\ast$ \\
\bottomrule
\end{tabular}

%% file: tables/tab_abl.tex
\begin{tabular}{lcc}
\toprule
Variant & DER & MAE \\
\midrule
\rowcolor{oursrow}Full recipe & 4.45 & 0.52 \\
always split (no capacity gate) & 4.70 & 0.89 \\
no slot splitting ($k_{\max}{=}1$) & 5.41 & 0.75 \\
no cannot-link merge guard & 4.45 & 0.55 \\
no embedding-confirmed rescue & 4.42 & 0.52 \\
no low-posterior rescue & 4.94 & 0.48 \\
one e-means restart & 4.41 & 0.52 \\
online-birth core, not official e-means & 4.43 & 0.50 \\
\bottomrule
\end{tabular}

%% file: tables/tab_ext.tex
\begin{tabular}{lccccccc}
\toprule
DER (\%) on 60 unseen sessions & $K{=}2$ & 4 & 6 & 8 & 10 & 12 & all \\
\midrule
Streaming Sortformer 4spk & 12.0 & 16.8 & 39.8 & 56.2 & 64.2 & 68.8 & 43.0 \\
Nemotron-3 head (8 slots) & 8.7 & 14.3 & 9.8 & 11.3 & 36.2 & 46.7 & 21.2 \\
Online Leader-Follower & 7.9 & 11.9 & 8.7 & 10.7 & 18.5 & 22.7 & 13.4 \\
\rowcolor{oursrow}Ours, as first emitted & 8.7 & 14.3 & 9.6 & 13.7 & 36.5 & 40.3 & 20.5 \\
\rowcolor{oursrow}Ours, after revision & 8.2 & 13.8 & 9.2 & 13.2 & 29.4 & 32.7 & 17.8 \\
\bottomrule
\end{tabular}

%% file: tables/tab_extboot.tex
\begin{tabular}{lc}
\toprule
Ours (after revision) $-$ competitor & $\Delta$DER [95\,\% CI] \\
\midrule
Streaming Sortformer 4spk & -25.2 [-29.5, -20.6]$^\ast$ \\
Nemotron-3 head (8 slots) & -3.4 [-5.0, -1.9]$^\ast$ \\
Online Leader-Follower & +4.4 [+2.6, +6.3]$^\ast$ \\
\bottomrule
\end{tabular}

%% file: tables/tab_ovsyn_level.tex
\begin{tabular}{lccccc}
\toprule
Overlap share of speech & 0\,\% & 11\,\% & 26\,\% & 39\,\% & all \\
\midrule
Streaming Sortformer & 60.1 & 58.9 & 59.0 & 60.0 & 59.5 \\
Nemotron-3 head & 25.5 & 31.8 & 39.9 & 42.7 & 35.0 \\
Leader--follower & 17.1 & 24.7 & 37.0 & 43.0 & 30.4 \\
\rowcolor{oursrow}Ours & 22.6 & 28.6 & 38.4 & 41.9 & 32.9 \\
\bottomrule
\end{tabular}

%% file: tables/tab_ovsyn_k.tex
\begin{tabular}{lccccc}
\toprule
Speakers $K$ & 6 & 8 & 10 & 12 & all \\
\midrule
Streaming Sortformer & 44.0 & 57.4 & 65.4 & 71.1 & 59.5 \\
Nemotron-3 head & 22.7 & 23.7 & 40.4 & 53.1 & 35.0 \\
Leader--follower & 21.8 & 22.8 & 36.6 & 40.5 & 30.4 \\
\rowcolor{oursrow}Ours & 22.3 & 25.9 & 37.4 & 45.9 & 32.9 \\
\bottomrule
\end{tabular}

%% file: tables/tab_overlap_split.tex
\begin{tabular}{lcccccc}
\toprule
 & \multicolumn{2}{c}{VoxConverse} & \multicolumn{2}{c}{AMI} & \multicolumn{2}{c}{ICSI} \\
overlap of the half (\% of speech) & 0.2 & 6.7 & 1.8 & 11.6 & 8.7 & 26.7 \\
 & low & high & low & high & low & high \\
\midrule
\multicolumn{7}{l}{\emph{streaming, 1.04\,s}} \\
Streaming Sortformer & 10.2 & 18.0 & 4.9 & 11.8 & 21.9 & 32.8 \\
Nemotron-3 head & 3.0 & 9.8 & 4.1 & 14.4 & 19.3 & 34.5 \\
Leader--follower & 3.4 & 8.3 & 4.9 & 16.4 & 21.5 & 40.4 \\
\rowcolor{oursrow}Ours & 2.8 & 8.7 & 4.2 & 14.1 & 19.4 & 34.4 \\
\midrule
\multicolumn{7}{l}{\emph{non-streaming}} \\
Sortformer 30.4\,s & 9.6 & 16.9 & 4.1 & 10.3 & 22.9 & 34.5 \\
Leader--follower & 4.2 & 8.3 & 4.6 & 15.3 & 20.2 & 38.9 \\
\rowcolor{oursrow}Ours & 1.7 & 7.3 & 3.9 & 13.9 & 19.0 & 33.8 \\
\midrule
\multicolumn{7}{l}{\emph{same inputs, different slot head}} \\
Sortformer head & 10.2 & 21.5 & 5.3 & 15.9 & 23.1 & 37.6 \\
\quad + our clustering & 3.3 & 9.9 & 4.4 & 19.8 & 21.3 & 39.2 \\
Nemotron-3 head & 3.1 & 9.7 & 5.5 & 13.9 & 18.4 & 33.7 \\
\quad + our clustering & 2.1 & 7.1 & 4.2 & 13.8 & 18.3 & 33.6 \\
\bottomrule
\end{tabular}

%% file: tables/tab_overlap.tex
\begin{tabular}{lcccccccc}
\toprule
 & & \multicolumn{2}{c}{Nemotron-3 head} & \multicolumn{2}{c}{Leader--follower} & \multicolumn{2}{c}{Ours} \\
Corpus & overlap (\% of speech) & DER & miss & DER & miss & DER & miss \\
\midrule
VoxConverse & 2.8 & 52.6 & 51.3 & 53.1 & 51.0 & 52.8 & 51.4 \\
AMI & 7.0 & 56.3 & 53.5 & 60.1 & 52.2 & 56.1 & 53.5 \\
ICSI & 18.9 & 60.7 & 58.1 & 62.2 & 57.6 & 60.5 & 58.1 \\
LibriSpeech-OpenSpk & 1.0 & 65.1 & 63.9 & 60.3 & 60.2 & 61.1 & 60.8 \\
\bottomrule
\end{tabular}

%% file: tables/tab_full_all.tex
\begin{tabular}{lcccccc}
\toprule
 & \multicolumn{2}{c}{VoxConverse, 8\,min} & \multicolumn{2}{c}{AMI, 34\,min} & \multicolumn{2}{c}{ICSI, 61\,min} \\
Full-length, streaming (1.04\,s) & DER & SCA & DER & SCA & DER & SCA \\
\midrule
Streaming Sortformer 4spk (1.04\,s) & 18.1 & 29.5 & 8.9 & 93.8 & 36.3 & 6.7 \\
Nemotron-3 head, 8 slots & 8.5 & 59.1 & 9.7 & 81.2 & 28.5 & 33.3 \\
Online Leader-Follower & 8.2 & 43.2 & 13.9 & 0.0 & 43.4 & 0.0 \\
\rowcolor{oursrow}Ours, after revision & 7.9 & 59.1 & 9.6 & 87.5 & 28.6 & 40.0 \\
\bottomrule
\end{tabular}

%% file: tables/tab_engines.tex
\begin{tabular}{lccccc}
\toprule
Clustering engine inside the splitting step & Vox & Libri & AMI & ICSI & ms / call \\
\midrule
e-means (official, 5 ep, 3 restarts) & 4.45 & 8.56 & 8.90 & 26.90 & 28.6 \\
e-means (1 restart) & 4.41 & 9.00 & 8.90 & 26.90 & 24.4 \\
Lloyd k-means (25 it, 3 restarts) & 4.43 & 8.92 & 8.90 & 26.90 & 7.2 \\
Lloyd k-means (5 it, 3 restarts) & 4.39 & 8.90 & 8.90 & 26.90 & 1.7 \\
mini-batch k-means (Sculley, 5 ep) & 4.37 & 8.82 & 8.90 & 26.90 & 13.9 \\
online-birth core (no k selection) & 4.43 & 9.71 & 8.90 & 26.90 & 0.0 \\
\bottomrule
\end{tabular}

%% file: tables/tab_scale.tex
\begin{tabular}{rrccccc}
\toprule
$n$ & $k$ & Lloyd 5 it (s / gap \%) & Lloyd 25 it (s) & e-means 1 ep & e-means 5 ep \\
\midrule
5{,}000 & 16 & 0.01 / +0.8 & 0.03 & 0.15 / +0.5 & 0.05 / -0.8 \\
5{,}000 & 128 & 0.02 / +1.1 & 0.03 & 0.03 / +2.4 & 0.05 / -2.7 \\
50{,}000 & 16 & 0.02 / +1.4 & 0.04 & 0.09 / -0.6 & 0.24 / -1.7 \\
50{,}000 & 128 & 0.11 / +1.5 & 0.42 & 0.08 / -0.8 & 0.27 / -2.1 \\
200{,}000 & 16 & 0.04 / +1.0 & 0.07 & 0.20 / -0.8 & 0.75 / -1.0 \\
200{,}000 & 128 & 0.13 / +1.1 & 0.44 & 0.20 / -1.1 & 0.93 / -2.4 \\
\bottomrule
\end{tabular}

%% file: tables/tab_engines_stream.tex
\begin{tabular}{lccccc}
\toprule
Engine in the streaming system & Vox & AMI & ICSI & Libri & ms / update \\
\midrule
e-means (default) & 5.8 & 9.1 & 27.4 & 16.4 & 49 \\
Lloyd $k$-means, 25 iterations & 5.8 & 9.1 & 27.4 & 16.3 & 38 \\
Lloyd $k$-means, 5 iterations & 5.8 & 9.1 & 27.4 & 16.1 & 36 \\
mini-batch $k$-means & 5.8 & 9.1 & 27.4 & 16.6 & 37 \\
\bottomrule
\end{tabular}

%% file: tables/tab_budget.tex
\begin{tabular}{llcc}
\toprule
Engine & Passes & Vox & Libri \\
\midrule
e-means & 1 & 4.34 & 8.56 \\
e-means & 2 & 4.39 & 8.67 \\
e-means & 3 & 4.45 & 8.56 \\
e-means & 5 & 4.45 & 8.56 \\
e-means & 10 & 4.45 & 8.87 \\
Lloyd & 1 & 4.55 & 9.93 \\
Lloyd & 2 & 4.41 & 9.60 \\
Lloyd & 3 & 4.38 & 9.11 \\
Lloyd & 5 & 4.39 & 8.90 \\
Lloyd & 10 & 4.44 & 8.86 \\
Lloyd & 25 & 4.43 & 8.92 \\
mini-batch & 1 & 4.34 & 9.68 \\
mini-batch & 2 & 4.34 & 9.36 \\
mini-batch & 3 & 4.36 & 9.23 \\
mini-batch & 5 & 4.37 & 8.82 \\
\bottomrule
\end{tabular}

%% file: tables/tab_gap.tex
\begin{tabular}{lcccc}
\toprule
Passes & e-means gap & Lloyd gap & e-means ms & Lloyd ms \\
\midrule
1 & 3.4 & 15.3 & 16.1 & 0.6 \\
2 & 2.1 & 7.6 & 17.0 & 0.7 \\
3 & 1.8 & 4.8 & 18.3 & 0.9 \\
5 & 1.6 & 3.4 & 21.0 & 1.3 \\
10 & -- & 2.5 & -- & 2.4 \\
25 & -- & 2.4 & -- & 5.3 \\
\bottomrule
\end{tabular}

%% file: tables/tab_sens.tex
\begin{tabular}{lcccc}
\toprule
Setting & Vox & AMI & ICSI & Libri \\
\midrule
default & 4.45 & 8.90 & 26.90 & 8.56 \\
act\_thr=0.45 & 4.38 & 9.01 & 27.36 & 8.53 \\
act\_thr=0.5 & 4.45 & 8.93 & 27.11 & 8.56 \\
act\_thr=0.55 & 4.45 & 8.90 & 26.90 & 8.56 \\
act\_thr=0.6 & 4.44 & 8.93 & 26.71 & 8.58 \\
act\_thr=0.65 & 4.46 & 9.01 & 26.48 & 8.60 \\
vad\_thr=0.3 & 4.44 & 8.86 & 27.13 & 8.33 \\
vad\_thr=0.4 & 4.45 & 8.90 & 26.90 & 8.56 \\
vad\_thr=0.5 & 4.50 & 8.94 & 26.79 & 8.89 \\
sat\_slots=6 & 4.30 & 8.90 & 28.62 & 8.28 \\
sat\_slots=7 & 4.28 & 8.90 & 27.56 & 8.56 \\
sat\_slots=8 & 4.45 & 8.90 & 26.90 & 8.56 \\
split\_cos=0.45 & 4.30 & 8.90 & 26.90 & 9.09 \\
split\_cos=0.6 & 4.45 & 8.90 & 26.90 & 8.56 \\
split\_cos=0.75 & 4.85 & 8.90 & 26.90 & 14.48 \\
min\_sub\_sec=1.0 & 4.45 & 8.90 & 26.90 & 8.56 \\
min\_sub\_sec=2.0 & 4.45 & 8.90 & 26.90 & 8.56 \\
min\_sub\_sec=4.0 & 4.45 & 8.90 & 26.90 & 8.56 \\
k\_max=2 & 4.71 & 8.90 & 26.90 & 10.44 \\
k\_max=3 & 4.51 & 8.90 & 26.90 & 8.50 \\
k\_max=4 & 4.45 & 8.90 & 26.90 & 8.56 \\
k\_max=6 & 4.46 & 8.90 & 26.90 & 8.56 \\
min\_spk\_sec=0.3 & 4.45 & 8.90 & 26.90 & 8.56 \\
min\_spk\_sec=1.0 & 4.45 & 8.90 & 26.90 & 8.56 \\
smooth\_sec=0.0 & 4.45 & 8.90 & 26.90 & 8.54 \\
smooth\_sec=0.4 & 4.45 & 8.90 & 26.90 & 8.56 \\
smooth\_sec=0.8 & 4.45 & 8.90 & 26.90 & 8.67 \\
\bottomrule
\end{tabular}

%% file: tables/tab_sabl.tex
\begin{tabular}{lcccc}
\toprule
Variant (DER after revision, \%) & Vox & AMI & ICSI & Libri \\
\midrule
Default (window 90\,s, revision 10\,s, centred correction) & 5.8 & 9.1 & 27.4 & 16.4 \\
window 30\,s & 6.1 & 9.1 & 27.4 & 18.6 \\
window 60\,s & 6.1 & 9.1 & 27.4 & 16.7 \\
window 180\,s & 5.5 & 9.1 & 27.3 & 16.2 \\
revision horizon 3\,s & 5.8 & 9.1 & 27.4 & 16.7 \\
revision horizon 30\,s & 5.8 & 9.1 & 27.4 & 16.2 \\
causal embeddings only & 5.8 & 9.2 & 27.5 & 17.4 \\
\bottomrule
\end{tabular}

%% file: tables/tab_aux.tex
\begin{tabular}{lcccccccc}
\toprule
 & \multicolumn{2}{c}{VoxConverse} & \multicolumn{2}{c}{AMI} & \multicolumn{2}{c}{ICSI} & \multicolumn{2}{c}{LibriSpeech} \\
System & MAE & cpWER & MAE & cpWER & MAE & cpWER & MAE & cpWER \\
\midrule
Streaming Sortformer 4spk, 30.4\,s config & 2.0 & -- & \textbf{0.2} & 14.9 & 1.5 & -- & 3.3 & 69.4 \\
Nemotron-3 (8 slots, 30.4\,s) & \underline{0.7} & -- & \underline{0.3} & \underline{13.2} & \textbf{0.5} & -- & 1.0 & 23.6 \\
pyannote 3.1 & 1.0 & -- & 0.9 & -- & \underline{1.3} & -- & 2.2 & -- \\
Online Leader-Follower & 2.5 & -- & 3.9 & 15.7 & 6.2 & -- & \underline{0.9} & \underline{9.8} \\
Mini-batch $k$-means (Sculley) & 1.7 & -- & 4.9 & 25.2 & 6.2 & -- & 3.1 & 19.4 \\
Lloyd $k$-means++ (25 ep.) & 1.3 & -- & 1.1 & 14.7 & 2.1 & -- & 1.7 & 38.5 \\
\rowcolor{oursrow}\textbf{Ours (non-streaming)} & \textbf{0.5} & -- & \underline{0.3} & \textbf{13.1} & \textbf{0.5} & -- & \textbf{0.1} & \textbf{6.2} \\
\midrule
Streaming Sortformer 4spk (1.04\,s) & 2.1 & -- & \textbf{0.2} & 16.6 & \underline{1.5} & -- & 3.4 & 71.9 \\
Nemotron-3 head, 8 slots (1.04\,s) & 0.7 & -- & \textbf{0.2} & \textbf{13.5} & \textbf{0.5} & -- & \textbf{1.0} & 25.5 \\
Online Leader-Follower (1.04\,s) & 2.3 & -- & \underline{4.8} & 16.8 & 6.7 & -- & \underline{1.4} & \textbf{12.5} \\
\rowcolor{oursrow}Ours, as first emitted (1.04\,s) & \underline{0.6} & -- & \textbf{0.2} & \underline{13.6} & \textbf{0.5} & -- & 2.1 & 24.2 \\
\rowcolor{oursrow}\textbf{Ours, after $\le$10\,s revision (1.04\,s)} & \textbf{0.5} & -- & \textbf{0.2} & \textbf{13.5} & \textbf{0.5} & -- & 1.8 & \underline{20.7} \\
\midrule
Nemotron-3 head, 8 slots (0.64\,s) & \underline{0.8} & -- & \textbf{0.2} & 13.3 & \textbf{0.5} & -- & \textbf{1.0} & 25.7 \\
Online Leader-Follower (0.64\,s) & 3.5 & -- & \underline{4.6} & 17.5 & \underline{7.5} & -- & \underline{1.3} & \textbf{12.7} \\
\rowcolor{oursrow}Ours, as first emitted (0.64\,s) & \textbf{0.6} & -- & \textbf{0.2} & \underline{13.2} & \textbf{0.5} & -- & 2.4 & 26.0 \\
\rowcolor{oursrow}\textbf{Ours, after $\le$10\,s revision (0.64\,s)} & \textbf{0.6} & -- & \textbf{0.2} & \textbf{13.1} & \textbf{0.5} & -- & 1.9 & \underline{21.1} \\
\bottomrule
\end{tabular}